# Contact angle hysteresis and oil film lubrication in electrowetting with two immiscible liquids

J. Gao,[1, a)] N. Mendel,[1, a)] R. Dey,[1, a)] D. Baratian,[1] and F. Mugele[1, b)]

[1]*Physics of Complex Fluids, Faculty of Science and Technology, University of Twente, P. O. Box 217, 7500AE, The Netherlands.*

[a] *These authors contributed equally to this work.*

[b] *Email: f.mugele@utwente.nl*

Electrowetting (EW) of water drops in ambient oil has found a wide range of applications including lab-on-a-chip devices, display screens, and variable focus lenses. The efficacy of all these applications is dependent on the contact angle hysteresis (CAH), which is generally reduced in the presence of ambient oil due to thin lubrication layers. While it is well-known that AC voltage reduces the effective contact angle hysteresis (CAH) for EW in ambient air, we demonstrate here that CAH for EW in ambient oil increases with increasing AC and DC voltage. Taking into account the disjoining pressure of the fluoropolymer-oil-water system, short range chemical interactions, viscous oil entrainment and electrostatic stresses, we find that this observation can be explained by progressive thinning of the oil layer underneath the drop with increasing voltage. This exposes the droplet to the roughness of the underlying solid and thereby increases hysteresis.

Electrowetting (EW) refers to the electrically enhanced wetting of a conductive sessile drop on a hydrophobic dielectric film, as quantified by the reduction in the apparent contact angle with the applied electrical voltage[1]. The reversible, reproducible and facile-to-implement nature of EW have established it as an efficient tool for active manipulation of discrete droplets in a broad range of applications. Most of these applications involve manipulation of water drops in an ambient oil, including many lab-on-a-chip devices, reflective displays, and optofluidic systems.[1] In addition to preventing evaporation, one of the key advantages of the in-oil configuration is the reduction of CAH and protection of the dielectric surface due to a thin oil film[2-5] under the droplet, which serves as a lubrication layer[4-6]. Minimum CAH ensures optimum droplet mobility[7,8] and reliability of operation. In case of EW in ambient air, Mugele and coworkers[9,10] demonstrated that CAH decreases with increasing AC voltage, while it remains essentially constant for DC voltage. The CAH reduction was attributed to depinning of the contact line from surface heterogeneity assisted by the oscillatory actuation of the contact line by local electric stresses. Resonance phenomena were shown to enhance this effect at low frequencies[11]. For the relevant practical operation in oil, however, the applicability of the same ideas is questionable. First of all, given the presence of the lubricating oil film, there isn't really a three phase contact line in many cases. Second, the lubrication film itself is subject to electrical stresses that can

– amongst others – lead to an electrohydrodynamic instability[3]. Third, the ambient medium is likely to dampen the oscillatory motion of the oil-water interface.

In the present letter, we demonstrate experimentally that the effect of EW on CAH in ambient oil is indeed strikingly different from the case of ambient air: in ambient oil CAH is found to *increase* with increasing voltage in the same manner for both AC and DC voltage. An analysis of molecular interaction forces and electrical stresses shows that this effect is caused by the fact that the oil film under the droplet decreases in thickness under the influence of the Maxwell stress and thereby gradually loses its ability to smoothen the intrinsic substrate roughness and to lubricate the contact line. The arguments presented here also explain why the use of liquid-infused porous substrates, also known as SLIPS surfaces, in EW[12] have so far been of limited success.

The experimental set-up is sketched in the inset of Fig. 1. For the CAH measurements, a glass substrate covered by a thin conductive layer of indium tin oxide (ITO) was used, which in turn was coated with a 2 μm thick insulating layer of parylene C and a top layer of hydrophobic Teflon with a thickness of around 50 nm. In order to observe the effect of surface roughness, a separate roughened substrate consisting of a thin layer of ITO and 2 μm thick Parylene C was prepared via oxygen-plasma treatment; subsequently, the roughened surface was coated with a conformal layer of fluorocarbon, which is chemically similar to Teflon, using chemical vapor deposition. Deionized water droplet (with added 0.1 M NaCl to increase conductivity) was deposited on the substrate in ambient oil (1-bromohexadecane, with a density of 0.99 g/cm$^3$ to avoid gravity affect, and a viscosity $\mu=8.51\times10^{-3}$ Pa·s; oil-water interfacial tension: $\gamma_{ow} = 42$ mN/m). DC or AC (frequency 10 kHz) voltages were applied between the ITO layer and a syringe needle which was immersed into the droplet (inset in Fig. 1a). For fixed applied voltage, the droplet was inflated or deflated using a syringe pump connected to the needle at a constant rate of 0.1 μl/s. During the inflation/deflation process the droplet volume varied between 10 μL to 25 μL, This leads to a contact line velocity of the order of $v \approx 10^{-6}$ m/s. Together with the oil viscosity and the interfacial tension, this leads to values of the non-dimensional velocity- the capillary number $Ca$ (based on the oil viscosity), of $Ca = \mu v/\gamma_{ow} < 10^{-6}$. For each experimental condition, a new droplet was placed on the substrate. After that, the voltage was slowly ramped up (~0.2 V/s) until the desired voltage was reached. The slow ramping prevents the abrupt formation of oil droplets under water[3] which may affect the CAH. Identical experimental procedures were followed for the different substrates.

Advancing and receding contact angles under DC-EW for different applied voltages on the untreated substrate and the roughened substrate are shown in Fig. 1. Upon inflating the droplet, the contact angle increases until the advancing contact angle ($\theta_{adv}$) is reached, whereas upon deflating the contact angle decreases until the receding contact angle ($\theta_{rec}$) is reached (Fig. 1). It is clearly visible that for both the substrates advancing and receding contact angles decrease with the increase of voltage, which is expected considering the EW equation[1], $\cos\theta = \cos\theta_Y + \eta$, where $\theta_Y$ is the Young's contact angle and $\eta = \epsilon_0\epsilon U^2/d\,\gamma_{ow}$ is the non-dimensional electrowetting number. $d$=2 µm is the thickness and $\varepsilon_d$=3.1 the dielectric constant of parylene C. Noticeably, the CAH ($\theta_{adv}$- $\theta_{rec}$) on the untreated substrate increases from about 2° at 0 V to about 7° at 30 V (compare Fig. 1a and c), as also mentioned qualitatively in a previous report [13]. Accordingly, $\Delta\cos\theta = \cos\theta_{rec} - \cos\theta_{adv}$ increases with increasing $\eta$ for both DC-EW and AC-EW in oil on the untreated substrate (Fig. 2a). For the roughened substrates, the CAH is found to increase even more strongly than for the untreated ones (Fig. 1 and Fig. 2a). In contrast, control experiments with water drops in ambient air on identical untreated substrate confirmed the earlier results[9,10] that the CAH gradually decreases over the same range of the applied AC voltage (Fig. 2b), while it remains approximately constant in case of DC voltage (Fig. S3 in the Supplemental Material). Fig. 2b clearly shows the opposing trends of CAH for ambient oil and ambient air. Furthermore, the CAH in oil is always smaller than that in air at low voltages, demonstrating the lubricating effect of the oil; however, at the high voltages investigated ($U$>35 V), CAH in oil and air merge at a value of $\Delta\cos\theta \approx 0.05$ for the present surfaces indicating a reduced ability to lubricate of the oil at higher voltage. (Measurements at higher voltage suggest that the hysteresis is approximately the same in air and oil in that regime. Yet, reproducibility is compromised and error bars are large, presumably due to the high local electric fields.)

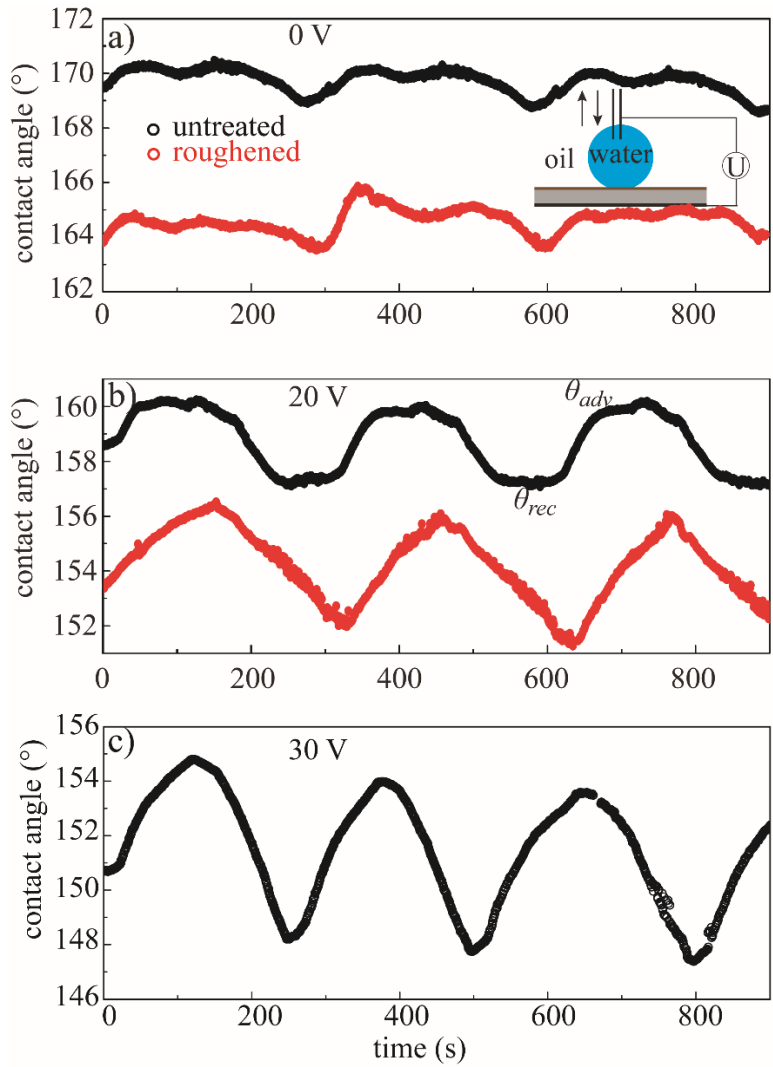

Fig. 1 Advancing and receding contact angles for a) 0 V ($\eta$=0), b) 20 V ($\eta$=0.065), and c) 30 V ($\eta$=0.147) under EW for untreated substrate (black) and roughened substrate (red). Note that higher than 20 V is not tested on the roughened substrate due to its comprised dielectric strength. Inset in a): simplified sketch of the set-up (not to scale).

To understand the increase in CAH for a water droplet under EW in oil, we first recollect that CAH originates from pinning of the contact line at topographical and/or chemical heterogeneities of the underlying solid surface[14,15]. The pinning strength of these defects is reduced if the surface is covered by a thin oil layer interspersed between the solid substrate and the drop. Such lubrication layers can arise for two reasons. First, they can be stabilized by molecular interaction forces. In this case, the films typically have an equilibrium thickness in the nanometer range[2,15]. Second, oil films can be dynamically entrained as the contact line spreads across the surface. In that case, the film thickness depends on the contact line velocity $v$ and scales as $Ca^{2/3}$, as described by the Landau-Levich model[15,16]. Depending on *Ca*, such entrainment films can reach much larger thicknesses up to the micrometer range[3,15]. Obviously, thicker films are much more efficient at screening heterogeneities on the surface than thinner ones: thick films are flattened by the oil-water interfacial tension and thereby smoothen any surface roughness and eliminate CAH, as illustrated in Fig. 3a. In contrast, thin films follow the contour of the surface and thereby expose the drop to the surface heterogeneity and cause CAH. The key idea of the model that we will present in the following is that the increase in CAH under EW in oil shown in Figs. 1 and 2 arises from the fact that the lubrication layer becomes thinner due to the increasing electrical stress upon increasing $U$, and thereby loses its ability to smoothen any surface roughness (Fig. 3a).

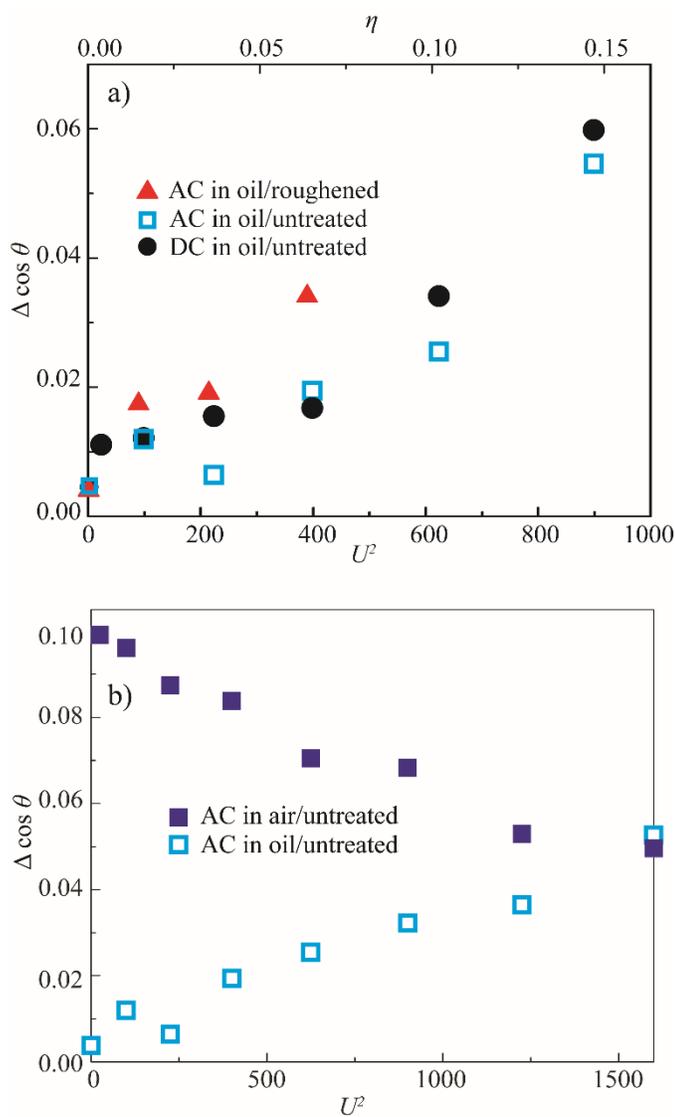

Fig. 2 $\Delta \cos \theta$ vs. $U^2$ for in-oil EW configuration (a), and comparison between $\Delta \cos \theta$ vs. $U^2$ for ambient air vs. ambient oil (b).

To demonstrate this, we first estimate the average thickness ($h$) of the oil film assuming, for the moment, that the underlying solid surface can be considered flat. This assumption is justified by the low aspect ratio of the surface roughness as characterized by Atomic Force Microscopy: the typical roughness amplitude of the untreated substrate is around 3 nm and the characteristic roughness wavelength of around or above 100 nm (Fig. 3b). The thickness of the static equilibrium film that is controlled by molecular interaction forces is determined by the minimum of the free energy (per unit area) $E(h)$ of the wetting film. Next to the interfacial energies of the substrate-oil interface $\gamma_{so}$ and $\gamma_{ow}$, it contains contributions from the electrostatic energy, the long range van der Waals interaction, and the repulsive short-range chemical interaction, which we can write as

$$E(h) = \gamma_{so} + \gamma_{ow} - \frac{\epsilon_0 \epsilon_d}{2d}\left(1 + \frac{\epsilon_d\, h}{\epsilon_{oil}\, d}\right)^{-1} U^2 + \frac{A}{12\pi h^2} + \varphi_0\, e^{-\frac{h}{\lambda}} \tag{1}$$

Here $A = -1.83 \times 10^{-21}$ J is the Hamaker constant for Teflon AF-oil-water as calculated based on the bulk refractive indices and dielectric constants using the standard expression[17]. With our sign convention, the negative sign of $A$ means that the van der Waals interaction for the Teflon-oil-water system is attractive, thus favouring partial oil wetting as it should be. (Note that this is different from the assumptions made in ref. 2, where $A$ was chosen to be positive in contrast to the typical situation in EW); $\varepsilon_{oil}$ is the dielectric constant of oil; $\lambda$ is the characteristic decay length for the short-range chemical interaction and should be in subnanometer scale. $\varphi_0 > 0$ is typically tens of mJ/m². Considering the realistic assumption that $h \ll d$, Eq. 2 can be rewritten as,

$$E(h) \approx E_{ref} + \eta\, \gamma_{ow} \frac{\epsilon_d\, h}{\epsilon_{oil}\, d} + \frac{A}{12\pi h^2} + \varphi_0\, e^{-h/\lambda} \tag{2}$$

where $E_{ref} = \gamma_{so} + \gamma_{ow}(1 - \eta)$ is the $h$-independent contribution. Fig. S1 in the supplementary material shows the shape of $E(h) - E_{ref}$ for typical values of $\lambda = 0.5$ nm, $\varphi_0 = 10$ mJ/m², and $\eta = 0\ldots 0.8$. Numerical minimization with respect to $h$ yields the static equilibrium thickness $h_{sta}$. It is found to decrease from about 4 nm at $\eta = 0$ to about 3 nm at $\eta = 0.8$ (Fig. 3c).

In case of a finite contact line velocity, we can estimate the thickness $h_{dyn}$ of the hydrodynamically entrained oil film using the adapted Landau-Levich model in the presence of electrowetting, as described in ref. 3. For this approach, it must be first realized that the entrapped oil film underneath the droplet (region I) and the moving droplet interface (region II) are bridged by a dynamical interface of length scale $l$ (Fig. 3a), which can be estimated by asymptotically matching the curvature of the dynamical interface with the macroscopic curvature of the spherical droplet ($1/R$), which leads to $l = \sqrt{hR}$. Within lubrication approximation, the pressure gradient across this dynamical region ($\Delta P/l$) is balanced by the viscous force ($\mu v/h^2$) within this region due to the flow field generated by the moving droplet interface. In this regard, $\Delta P$ can be estimated by the difference between the pressures at regions I and II (Fig. 3a), where the pressure at region I consists of the contributions due to electrostatic interaction, van der Waals interaction, and chemical interaction (obtained by taking the derivative of $E$, as given by Eq. 1) while the pressure at region II consists only of the Laplace pressure. Returning to the full expression for the electrical stress (because the assumption $h \ll d$ no longer holds), we can write the force balance equation that determines the values of $h$ as

$$\eta \frac{\varepsilon_d}{\varepsilon_{oil} d}\left(1+\frac{\varepsilon_d h}{\varepsilon_{oil} d}\right)^{-2} - \frac{A}{\gamma_{ow} 6\pi h^3} - \frac{\varphi_0}{\gamma_{ow} \lambda} e^{-h/\lambda} + \frac{2}{R} \approx Ca \frac{R^{1/2}}{h^{3/2}} \tag{3}$$

Typically, in our experiments, $Ca \approx 2\times10^{-7}$, $R \approx 10^{-3}$ m. For $\eta = 0\ldots 0.8$, $h_{dyn}$ can be calculated by numerically solving Eq. 3, and as shown in Fig. 3c, $h_{dyn}$ decreases with increasing $\eta$. For $\eta > 0.05$, however, $h_{dyn}$ and $h_{sta}$ almost overlap, which means that $h_{dyn}$ is mainly determined by the electrostatic pressure, the van der Waals interaction, and the short-range chemical interactions for small $Ca$. It is also worth noting that for the thicknesses estimated here, the oil film is stable; i.e. it does not suffer from the electrohydrodynamic instability that leads to the formation of oil drops in many practical EW experiments. This can be concluded from a linear stability analysis as in ref. 3 (see Fig. S2). In essence, the variations of $h_{dyn}$ and $h_{sta}$ for small $Ca$, as shown in Fig. 3c, suggest that hydrodynamic entrainment plays a minor role for the EW-induced increase in CAH in the present experiments.

Comparison of Fig. 3b and Fig. 3c shows that the estimated oil film thickness is comparable to the roughness amplitude of the underlying untreated hydrophobic substrate. This raises the question to what extent the oil film can smooth out the substrate roughness. From a balance of the molecular interaction forces described in Eq. 3 and the surface tension forces that tend to smooth out the oil-water interface, we can define a healing length $\xi_h$ [18], which characterizes the transition from replicating the substrate topography for lateral roughness scales $\zeta > \xi_h$ to smoothing out roughness on shorter length scales ($\zeta < \xi_h$):

$$\xi_h \approx \sqrt{\frac{\gamma_{ow}}{\left|\frac{\partial^2 E}{\partial h^2}\right|}} = \sqrt{\frac{1}{\left|\frac{1}{\gamma_{ow}}\frac{A}{2\pi h^4} - 2\eta \frac{\varepsilon_d^2}{\varepsilon_{oil}^2 d^2}\left(1+\frac{\varepsilon_d h}{\varepsilon_{oil} d}\right)^{-3} + \frac{\varphi_0}{\gamma_{ow}\lambda^2} e^{-h/\lambda}\right|}} \tag{4}$$

Using both $h_{sta}$ and $h_{dyn}$ from above, we find that $\xi_h$ decreases in both cases with increasing $\eta$, as expected. For the specific conditions of our present experiments ($Ca \approx 2\times10^{-7}$; blue curve in Fig. 3d), $\xi_h$ is found to decrease from values of $\approx 100\mu m$ at zero voltage ($\eta = 0$), i.e. much longer than the characteristic wavelength $\zeta$ of the roughness of the untreated substrate shown in Fig. 3b, to values $< 100nm$ for higher voltages. The calculation thus confirms our hypothesis: at zero voltage, the surface roughness is smoothed out by the dynamically entrained thick oil film; for finite voltage, the average thickness of the entrained film decreases and hence healing length drops below the characteristic scale of the substrate roughness thus exposing the drop to the heterogeneity and increasing contact line pinning and CAH. Based on this discussion, it is also expected that the CAH at a finite voltage should further increase if the surface roughness amplitude is increased, while at zero voltage it should remain largely unchanged. This is exactly what we observe for the CAH measurements on roughened substrate (with a roughness amplitude of ~20-60 nm (Fig. 3b)): The resulting CAH on the roughened substrate (triangles in Fig. 2a) is indeed higher than that of the non-treated substrates (squares in Fig. 2a) at finite AC voltages, while the CAH is almost the same for the two substrates at zero voltage (see Fig. 2a). Such variations of CAH under EW in oil, for substrates of varying roughness, is qualitatively in accordance with our hypothesis.

In many practical applications of EW, $Ca$ changes over many orders of magnitude. As the voltage is abruptly switched to actuate a drop, $Ca$ can reach values from $10^{-5}$ to $10^{-3}$ [3]. For such high $Ca$, hydrodynamic entrainment is significantly enhanced and $h_{dyn}$ reaches values up to micrometers. For instance, at $Ca = 2\times10^{-5}$, $h_{dyn}$ (open yellow circles in Fig. 3c) and $\xi_{h\text{-}dyn}$ (open yellow circles in Fig. 3d) are both orders of magnitude larger than the roughness amplitude and the roughness wavelength, respectively. For a chosen $\eta = 0.2$, as $Ca$ increases, $h_{dyn}$ gradually deviates from $h_{sta}$ (solid line in the inset of Fig. 3c). At large $Ca$, hydrodynamic entrainment dominates. This leads to $h_{dyn} \sim Ca^{2/3}$ (dashed line in the inset of Fig. 3c). Under such dynamic conditions, the surface roughness is thus expected to be efficiently screened by the thick lubrication layer, in agreement with the large number of practical observations indicating facile contact line motion for EW in ambient oil. When the voltage is kept fixed, however, $h$ is always expected to relax back to its static equilibrium value involving higher hysteresis.

In summary, our analysis establishes a relationship between voltage, substrate roughness, contact line velocity and CAH for EW experiments in ambient oil. In line with practical experience, high voltage and longer lateral roughness scales are found to reduce the lubricating efficiency of the oil and thus to enhance contact angle hysteresis in the static limit. For capillary numbers of $\approx 10^{-6}$ and higher, a simple lubrication model suggests that hydrodynamic entrainment progressively alleviates the effect of the voltage by enabling thicker much hydrodynamically entrained oil layers. A more quantitative analysis of the coupling between the roughness profile and CAH is conceivable. Yet, this will require a complete hydrodynamic analysis of the problem, which is beyond the scope of the present study.

**Supplementary Material**

See supplementary materials for results of $E$-$E_{ref}$ vs $h$ for $\eta = 0\ldots0.8$ (Fig. S1), stability analysis (Fig. S2), and $\Delta \cos \theta$ vs. $U^2$ for in-air DC-EW configuration (Fig. S3).

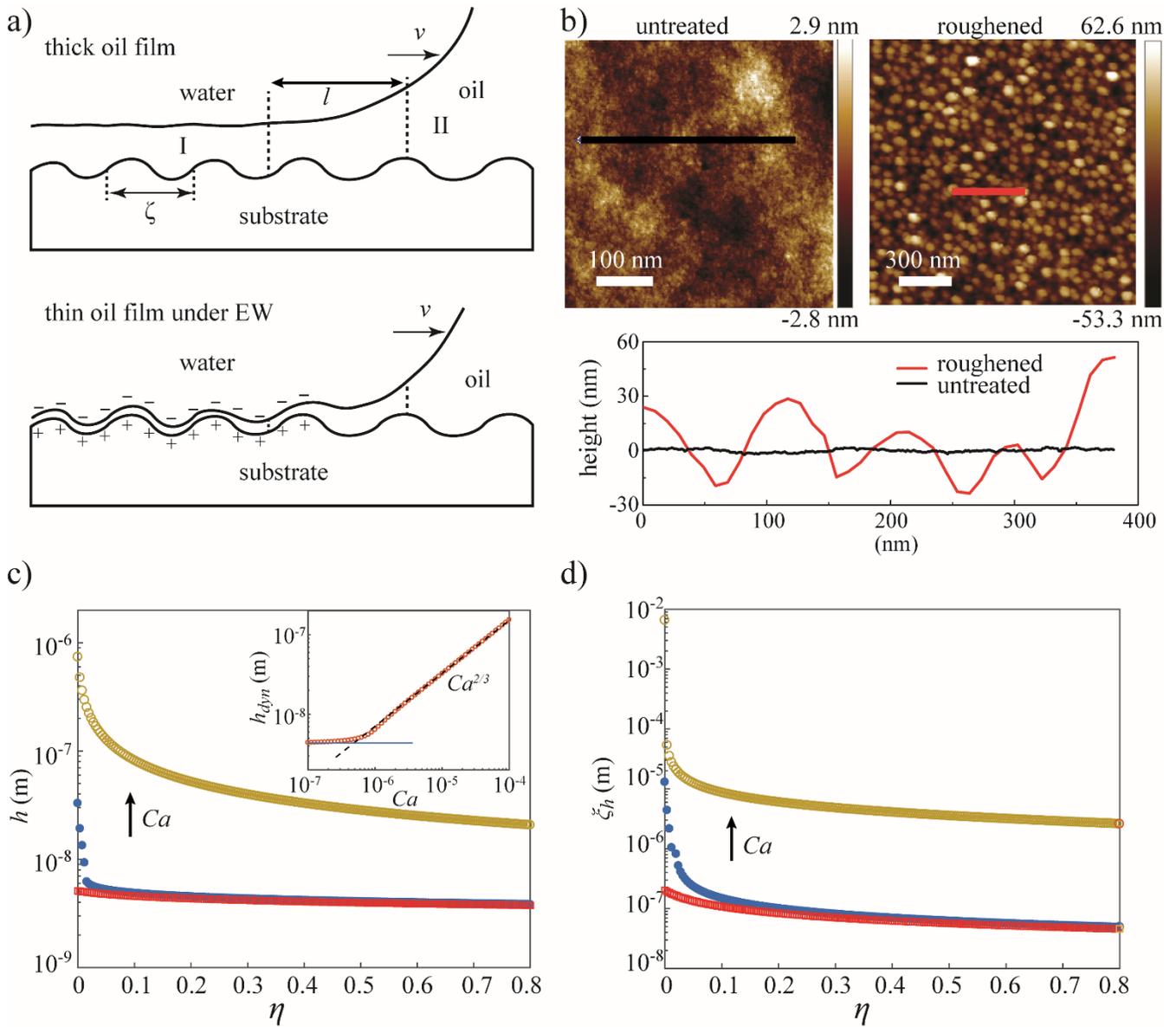

Fig. 3 Analysis of mechanism. a) Sketch of the oil film entrained between water droplet and substrate, which follows the surface roughness profile. b) AFM image of the untreated and roughened substrate. c) $h_{dyn}$ at $Ca=2\times10^{-5}$ (yellow open circles) and $Ca=2\times10^{-7}$ (blue solid circles), and $h_{sta}$ (red open squares) for $\eta =0…0.8$. Inset: $h_{dyn}$ for $Ca=10^{-7}…10^{-5}$ at $\eta =0.2$. d) $\xi_{h\text{-}dyn}$ at $Ca=2\times10^{-5}$ (yellow open circles) and $Ca=2\times10^{-7}$ (blue solid circles), and $\xi_{h\text{-}sta}$ (red open squares) for $\eta =0…0.8$.

**Supplementary Material**

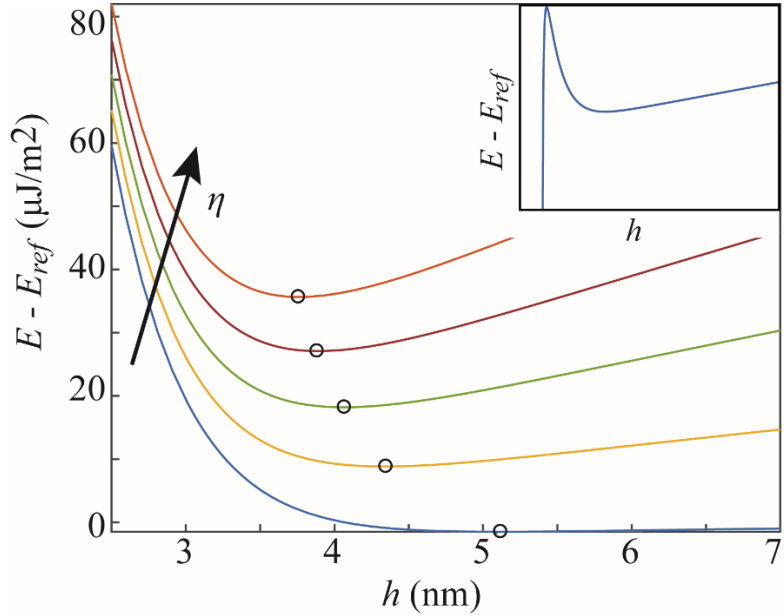

Fig. S1 $E$-$E_{ref}$ vs $h$ for $\eta$ = 0, 0.2, 0.4, 0.6, 0.8 (from bottom to top) at $h$ = 2.7 nm. Open circles indicate the position of the minima. Inset sketches the global trend of $E$-$E_{ref}$ vs $h$.

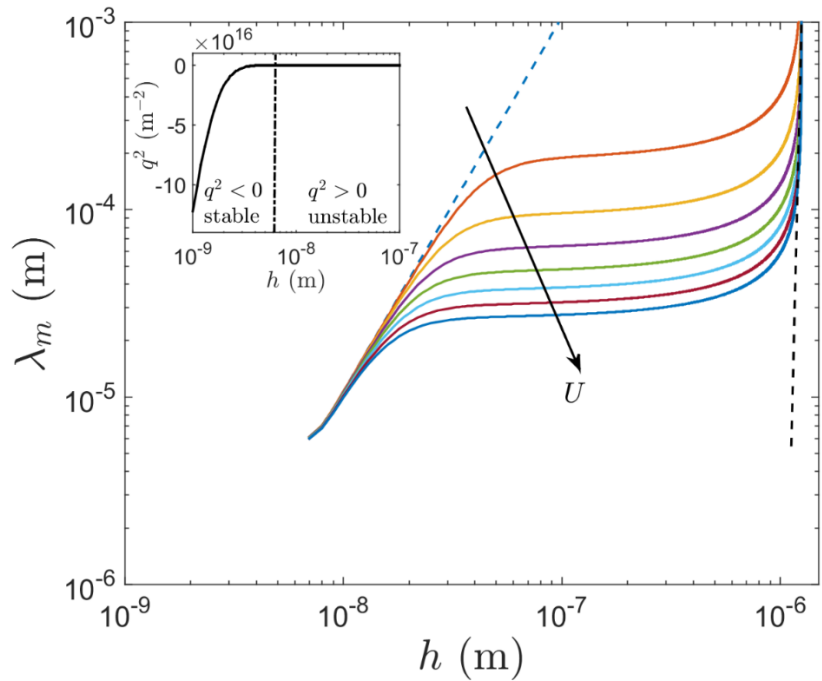

Fig. S2 Variations of the fastest growing unstable wavelength with the thickness of the oil film for different values of the applied electrical voltage. The thin film instability analysis has been performed considering the electrostatic pressure, and the contributions of the van der Waals and the short range chemical

interactions to the local pressure. This analysis shows that for $h \sim O(1) - O(10)$ nm the film is unconditionally stable (the inset shows that there is no solution for the wavenumber over that range of the film thickness). Even for the unstable films, the fastest growing wavelengths are of the order of tens of microns which are significantly larger than the roughness length scale present in the system.

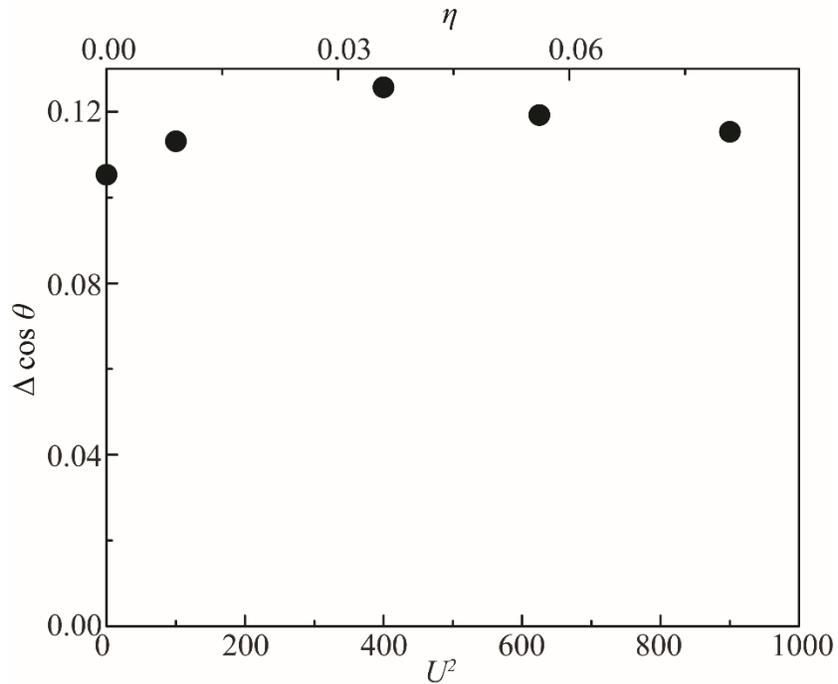

Fig. S3 $\Delta \cos \theta$ vs $U^2$ and $\eta$ for in-air DC EW configuration. CAH remains approximately constant in case of DC-EW in air